\documentclass[a4paper,11pt]{article}
\usepackage{amsmath,amssymb,mathtools,cite,graphicx,fancyhdr}
\usepackage{bbold,color,mathpazo,authblk}
\usepackage{caption}
\graphicspath{{Figures/}}
\newcommand{\dd}{\text{d}}

\providecommand{\abs}[1]{\lvert#1\rvert}

\usepackage{hyperref}
\hypersetup{
	unicode=false,          
	pdftitle={Hawking radiation in optics and beyond},    
	pdfauthor={David Bermudez},     
	pdfsubject={Analogue Gravity},   
	pdfcreator={David Bermudez},   
	pdfproducer={David Bermudez}, 
	pdfkeywords={analogue gravity, Hawking radiation, optical fibres}, 
	pdfnewwindow=true,      
	colorlinks=true,       	
	linkcolor=magenta,      
	citecolor=blue, 			
	filecolor=red,      		
	urlcolor=red           	
}

\begin{document}
\lhead{Hawking radiation in optics and beyond}
\rhead{Aguero-Santacruz, Bermudez}

\title{Hawking radiation in optics and beyond}
\author{Raul Aguero-Santacruz\footnote{{\it email:} raguero@fis.cinvestav.mx} \ and David Bermudez\footnote{{\it email:} dbermudez@fis.cinvestav.mx}}
\affil{\textit{Departamento de F\'{\i}sica, Cinvestav, A.P. 14-740, 07000 Ciudad de M\'exico, Mexico}}

\renewcommand\Authands{, and }

\date{}
\maketitle

\begin{abstract}
Hawking radiation was originally proposed in astrophysics, but it has been generalized and extended to other physical systems receiving the name of analogue Hawking radiation. In the last two decades, several attempts have been made to measure it in a laboratory, one of the most successful systems is in optics. Light interacting in a dielectric material causes an analogue Hawking effect, in fact, its stimulated version has already been detected and the search for the spontaneous signal is currently ongoing. We briefly review the general derivation of Hawking radiation, then we focus on the optical analogue and present some novel numerical results. Finally, we call for a generalization of the term Hawking radiation.\\

\end{abstract}

\section{Introduction}

The fascinating phenomenon of Hawking radiation was proposed over 45 years ago by Stephen W. Hawking \cite{Hawking1974}. Using a semi-classical approximation, i.e., considering quantum fields around a classical geometry, Hawking realized that the event horizon of a black hole in the Schwarzschild metric \cite{Misner2017} should emit massless particles. Black holes are not entirely black.

This is a significant phenomenon because it gives black holes---known as the ultimate devourers---something unthinkable,  a mechanism to lose mass. This black hole evaporation could dramatically change the cosmological evolution and the final state of the universe.

However, astrophysical Hawking radiation is too small to be directly detected by any conceivable experiment, satellite, or telescope up to now. This is partly because gravity is the weakest known force \cite{Feynman2006}. For example, in the Millikan oil drop experiment, a single electron charge can counteract the gravitational attraction of the whole planet earth. A quantum effect due to gravity would be extremely hard to measure.

This situation did not stop the theoretical studies of the effect. In 1981, while William Unruh was teaching a course on fluid dynamics \cite{Unruh2020}, he realized that the equations for the propagation of sonic waves in a moving fluid could be expressed in a similar fashion as those for light waves around the black-hole geometry as studied by Hawking \cite{Unruh1981}. This work slowly inspired new ideas on how to study phenomena usually related to gravity in other physical systems, giving birth to a new research area now called analogue gravity.

Inversely, general relativity inspired a connection between spacetime geometries and optical media through what became known as transformation optics \cite{Leonhardt2006,Pendry2006}, using the tools developed by Walter Gordon \cite{Gordon1923} and Jerzy Pleba\'{n}ski \cite{Plebanski1960} to study electromagnetic fields in curved spacetimes. Furthermore, transformation optics inspired a system made of light pulses interacting inside optical fibres as another scenario for analogue Hawking radiation \cite{Philbin2008,Bermudez2016,Jacquet2018,Drori2019}.

Note that analogue gravity does not reproduce the dynamics of general relativity given by Einstein equations, but rather, the kinematic of its solutions given by spacetime metrics: Analogue gravity can reproduce only the kinematic phenomena due to gravity and not the dynamical ones. Luckily, there are several kinematic effects to analyse, such as Hawking radiation \cite{Visser2003}.

To study the regime of ``semi-classical'' gravity, the field propagating on top of a curved background should be quantized without changing the energy-momentum tensor \cite{Birrell1982}. Unruh suggested liquid Helium for the first quantum analogue in 1981 \cite{Unruh1981}---see also work by Grigory Volovik \cite{Volovik2003}. Since then, several systems have been proposed, including phonons in Bose-Einstein condensates \cite{Garay2000,Munoz2019}, polaritons \cite{Nguyen2015}, and photons in optical fibres \cite{Philbin2008,Belgiorno2010,Drori2019}. Classical systems such as surface waves in water tanks have also been explored \cite{Weinfurtner2011,Euve2016}.

Research on gravity-like effects in analogue systems can give new clues into how to quantise gravity in some emergent gravity models by analysing the corresponding underlying theory in the analogue system, e.g., by investigating the role of high-frequency modes \cite{Unruh1995}, quantum vacuum, or dispersion in the derivation of Hawking radiation. One open question is still if astrophysical black holes radiate \cite{Helfer2003}. Furthermore, there are several interesting phenomena to study in analogue gravity besides Hawking radiation, such as cosmological creation of particles \cite{Fedichev2004}, dark energy \cite{Leonhardt2019}, and superradiance \cite{Torres2017}.

In this work we present briefly the general theory of analogue gravity, for more complete reviews see Refs. \cite{Bermudez2016,Unruh2007,Barcelo2011,Faccio2013,Belgiorno2019,Barcelo2019}. Then, we focus on the optical analogue and in particular on its theoretical and numerical aspects. For a review on the experimental side see the work of Yuval Rosenberg in this same special issue \cite{Rosenberg2020}. We explain how the Hawking effect works in the optical analogue and what are the particularities of this system, such as the strong dispersive regime and the completely non-thermal Hawking spectrum \cite{Leonhardt2012,Bermudez2016pra,Linder2016,MorenoRuiz2019}.

In Section 2, we review important aspects of astrophysical Hawking radiation and the kinematics of horizons. Section 3 is a summary of the theory of analogue systems describing modes around the horizon. In Section 4, we translate the developed tools to the spontaneous effect of the optical analogue. We describe the dynamics of the stimulated optical analogue with classical electromagnetism in Section 5. At last, we present our conclusions in Section 6.

Finally, given some resistance from a part of the scientific community, it is important to argue in favour of using the name Hawking radiation to describe this phenomenon in analogue systems. Here we argue why the name is appropriate and call for a generalization of the term Hawking radiation to include the same effect in systems different from astrophysics.

\section{Astrophysical Hawking radiation}
In the first half of the 1970s, Jacob Bekenstein's ideas about black-hole thermodynamics \cite{Bekenstein1973} were spreading and gain even more relevance once Hawking deduced his famous relation\footnote{In fact, Hawking asked for this equation to be written on his grave stone on Westminster Abbey, UK.} between the mass and the temperature of a radiating black hole \cite{Hawking1974}:
\begin{equation}
 T=\frac{\hbar c^3}{8\pi G M k_B}.
\end{equation}
This deduction works for black holes formed by gravitational collapse and whose event horizon is an apparent horizon in view of its finite creation time. If the mass of the star is larger than both the Chandrasekar and the Tolman-Oppenheimer-Volkoff (TOV) limits, the dying star cannot counterbalance its own weight and it collapses into a black hole.

Some of Bekenstein's equations relate the entropy with the surface area of a black hole and resemble the usual laws of thermodynamics. Bekenstein's deductions came from the dynamics of general relativity---given by Einstein equations---whereas the Hawking formula comes from the Schwarzschild metric and it only needs the kinematics of general relativity \cite{Visser2003}. Although the connection seems strong between both works, a true link is yet to be proven.

On its own, Hawking's deduction has some issues that have not been solved yet. The backreaction of the radiation field on the background is neglected. Being a semi-classical approach, quantum gravity effects are also neglected and the nature of pair production is not entirely clear. For an observer far away from the horizon, the measured waves are red-shifted, this means that waves are blue-shifted without limit the closer they are to the horizon, quickly getting even below the Planck length. This is known as the trans-Planckian problem \cite{Helfer2003}.

The study of Hawking radiation in analogue systems has already help us understand different aspects of this phenomenon. For example, it has been proven that if we introduce a cut-off in high-frequency modes,  the Hawking effect still persists \cite{Jacobson1991}. These modifications usually lead to grey factors that change the thermality of the spectrum.

Another important factor is the possibility to measure the Hawking effect offered by analogue experiments . If astrophysical black holes were to radiate, their temperature would be of the order of $\sim 10^{-8}$ K. Considering that the cosmic microwave background (CMB) temperature is $\sim 2.7$ K and its fluctuations of $\sim 10^{-4}$ K, we lack technological capacity to perform a direct measurement; thus making analogues the most viable alternative in search of experimental evidence for this effect.

The basic assumptions to derive Hawking radiation in any system were worked out by Matt Visser \cite{Visser2003}. The essential feature is the existence of a metric with the following characteristics: (1) apparent horizon, (2) non-zero surface gravity, and (3) slow evolution. We show that these points are fulfilled for the optical analogue. No dynamical considerations are used in this derivation: Bekenstein's theory of black-hole thermodynamics is not necessary.

\begin{figure}\centering
	\includegraphics[width=0.9\textwidth]{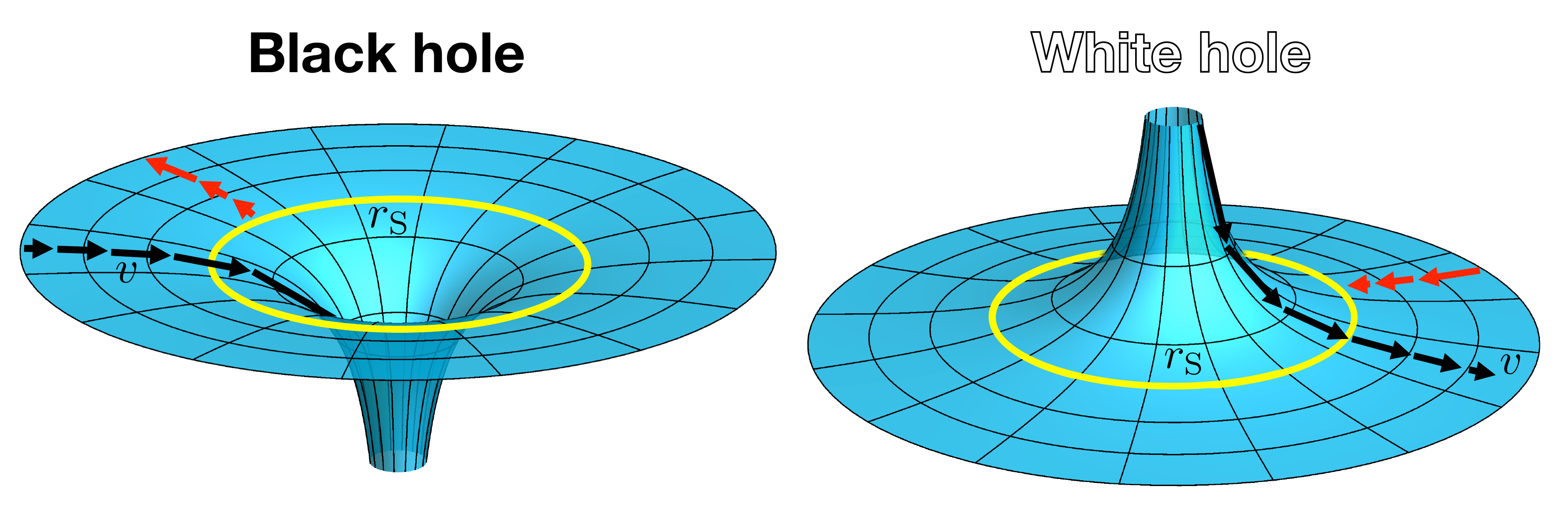}
	\caption{Embedding diagram of a black hole and a white hole. Waves travelling against the curvature (red) can be seen as counter-propagating waves in a fluid flow (black). The event horizon is marked with $r_\text{S}$ (yellow).}
	\label{fig_spacefluid}
\end{figure}

To elaborate on how the analogues work, we can picture Schwarzschild spacetime as a moving fluid for both of its solutions, the black and white holes. For a black hole, spacetime can be seen as being dragged into the centre of the hole; for the white hole---which is nothing more than a black hole evolving backwards in time---spacetime is being expelled out from the centre. For each case, waves travelling against the curvature are counter-propagating to the spacetime fluid. This is depicted in the embedding diagram in Fig. \ref{fig_spacefluid}.

\begin{figure}\centering
\includegraphics[width=0.45\textwidth]{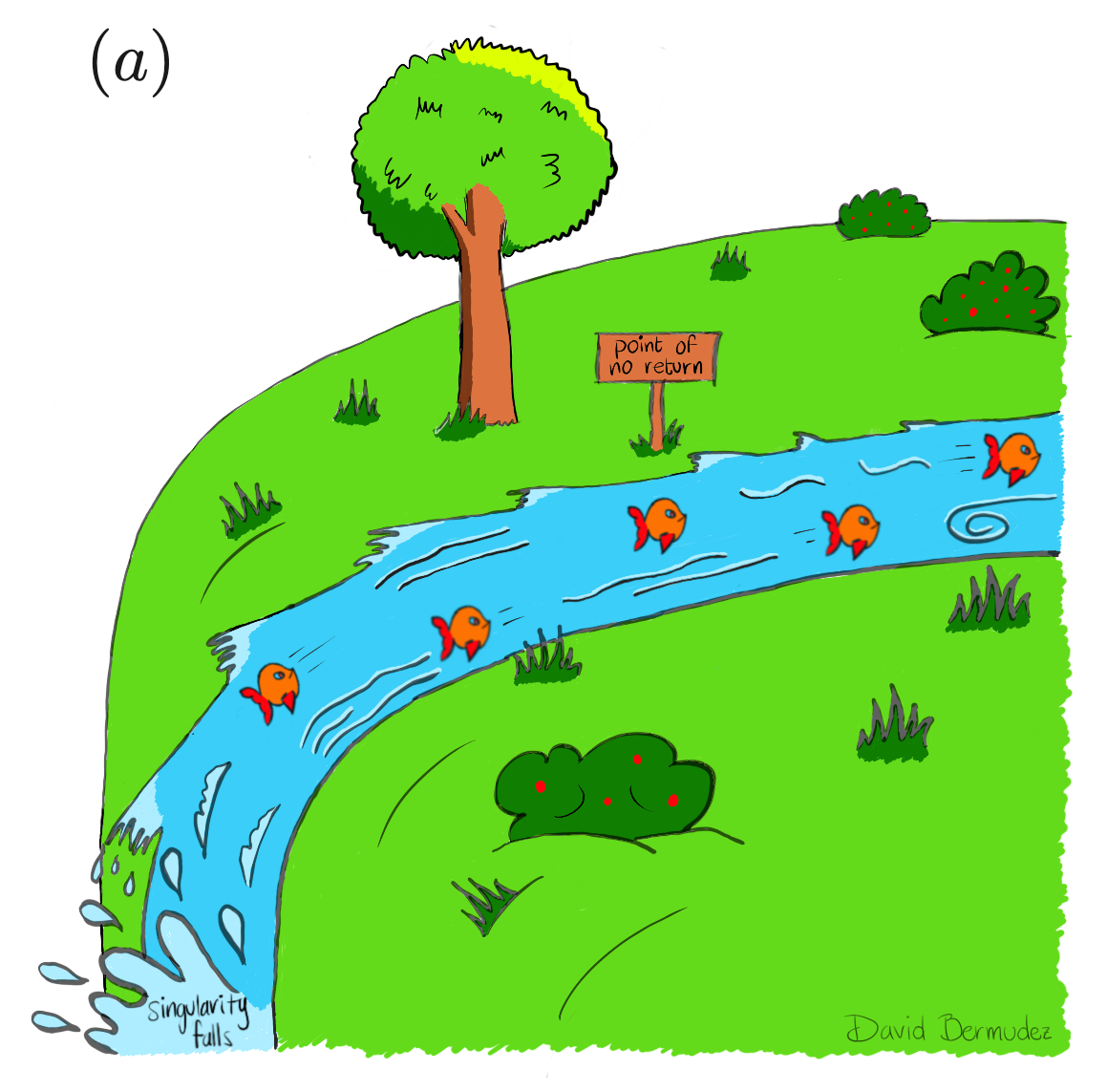}
\includegraphics[width=0.4\textwidth]{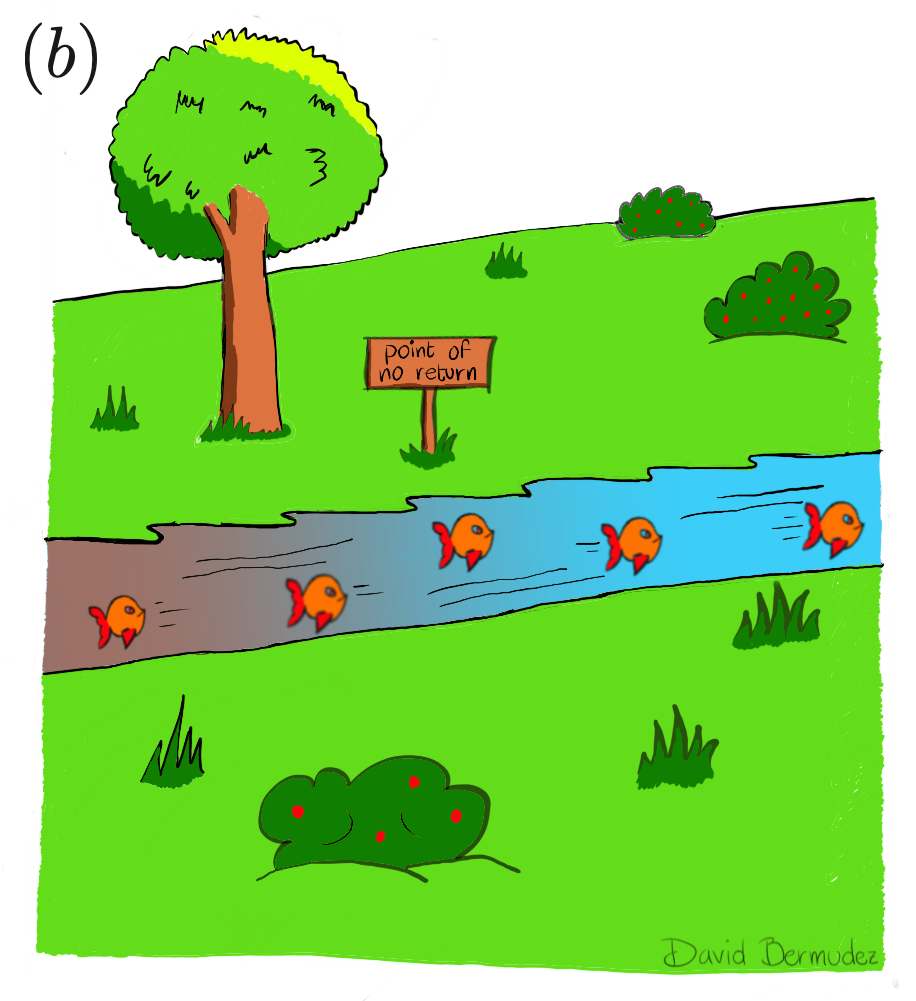}
\caption{River flow $V$ and fish with speed $c$. In (a), the river flow depends on the position $V=V(x)$ as its slope changes towards a waterfall, this is the situation in condensed-matter analogues. In (b), the fish speed changes with position $c=c(x)$ as water becomes muddier, this is the situation in optical analogues.}
\label{fig_fish}
\end{figure}

One of the main ingredients of Hawking radiation is an effective geometry that is divided by a horizon. We can picture this idea in a more familiar scenario using an analogy inspired by Unruh: a fish swimming in a river \cite{Unruh1981}. The river is considered as the background medium with an effective geometry and a flux speed $V$. The fish represents a perturbation propagating in this background, assuming it can only swim up to a maximum speed $c$, its destiny will depend on its relation with $V$. The river flow speed can change with position $V(x)$---e.g., increasing towards a waterfall (see Fig. \ref{fig_fish}(a))---or in a similar case, the fish speed can depend on position $c(x)$---e.g., if the water is becoming muddier and fish find it more difficult to swim there (see Fig. \ref{fig_fish}(b)). In a region where $V<c$, the fish will be able to swim in counter-current in an ever increasing speed up to a maximum of $c$ (right-hand side in both cases of Fig. \ref{fig_fish}). In the opposite case, where $V>c$, the fish is dragged by the current (left-hand side in both cases of Fig. \ref{fig_fish}). The horizon then becomes the point where $V=c$, the only place where the fish will stay put (marked with the sign ``point of no return''). We can summarize the important aspects as follows: (1) fluid flow $V$, (2) fish speed $c$, and (3) change of fluid flow $V=V(x)$ or of fish speed $c=c(x)$. Experiments in condensed-matter analogues change the fluid flow $V(x)$, while in optical analogues change the wave speed $c(x)$.

\section{General theory of analogue gravity}\label{general}
We associate the idea of a metric with coordinates in astrophysics, however, it can also describe effective geometries in a wide variety of physical scenarios, such as fluid flow or dielectric media configurations. Hawking radiation was first discovered in astrophysics using the Schwarzschild metric, which in Painlev\'e-Lema\^itre-Gullstrand (PLG) coordinates is
\begin{equation}
\dd s^2=-c^2\dd t^2+\left(\dd r+\sqrt{\frac{r_\text{S}}{r}}\,c\,\dd t \right)^2+r^2\dd \Omega^2,
\end{equation}
where $r_\text{S}$ is the Schwarzschild radius. We can generalize this metric to other velocity profiles $c(r,t)$ and $V(r,t)$ as
\begin{equation}
\dd s^2= -c(r,t)^2 \dd t^2 +[\dd r-V(r,t)\dd t]^2 +r^2 \dd \Omega^2,
\end{equation}
or equivalently as
\begin{equation}\label{PGmetric}
\dd s^2 = -[c(r,t)^2-V(r,t)^2]\dd t^2-2V(r,t)\dd r \dd t+\dd r^2+r^2 \dd\Omega^2,
\end{equation}
where $V(r,t)$ represents the velocity of a propagating medium and $c(r,t)$ the velocity of a perturbation propagating through such medium.

The analysis is usually performed by considering only one spatial dimension by ignoring the angular part.
In the matrix representation, the PLG metric takes the form
\begin{equation}
g_{\mu\nu}(\textbf{r},t) =\begin{pmatrix} -(c^2-V^2) & -V\hat{r}_j \\ -V\hat{r}_i & \delta_{ij}\end{pmatrix}.
\end{equation}
We can return to the Schwarzschild solution setting $c(r,t)=c$ the speed of light and $V(r,t)=-c\sqrt{r/r_\text{S}}$ the velocity profile of an observer in a radial trajectory. Due to radial symmetry, the horizon is found at the spatial region where the condition $c(r,t)=\abs{V(r,t)}$ is fulfilled. This could be an extended region or one or more points and it is not a physical singularity since $\text{det}(g_{\mu\nu})=-c(r,t)^2$. It is convenient to define the quantity $g_H(t)$ evaluated at the horizon as
\begin{equation}
g_H(t) =\left.\frac{1}{2} \frac{\dd [c(r,t)^2-V(r,t)^2]}{\dd r} \right| _H
		 =\left. c_H \frac{\dd [c(r,t)-\abs{V(r,t)}]}{\dd r} \right| _H,
\end{equation}
where $c_H$ is $c$ at the horizon. For the static geometry case, the definition of surface gravity $\alpha$ can be recovered \cite{Visser1993,Visser2003}:
\begin{equation} \label{kappa}
\alpha = \frac{g_H}{c_H}.
\end{equation}

The procedure to derive the Hawking temperature is the same as in the astrophysical case: by considering a free massless scalar quantum field $\phi(r,t)$ over a classical spacetime background with an associated curvature. This is why Hawking radiation is a semi-classical effect. Taking this field over the PLG spacetime in $1+1$ dimensions, i.e., ignoring the angular part $\dd\Omega =0$ and $r\rightarrow x$, the field Lagrangian can be expressed as \cite{Mukhanov2007}:
\begin{equation}\label{lag}
L = \frac{1}{2} \sqrt{-g} g^{\mu\nu} \partial_\mu \phi^\dag \partial_\nu \phi.
\end{equation}
When the analysis is done in 3+1 dimensions we arrive at a similar result: Hawking radiation is still emitted but its spectrum is not thermal, since it is modified by a grey-body factor \cite{Mukhanov2007}.
If we substitute the metric \eqref{PGmetric} and use the variational method or plug in the result in the Euler-Lagrange equation, we arrive to the wave equation
\begin{equation} \label{waveeq}
(\partial_t+\partial_x V)(\partial_t+V\partial_x )\phi -c^2\partial^2_x \phi =0.
\end{equation}

At this point we could call upon the eikonal approximation for the $s$ wave
\begin{equation}\label{eikonal}
\phi(x,t)=\mathcal{A}(x,t)\exp\left[\mp i\left(\omega t -\int^x k(y)\dd y \right) \right],
\end{equation}
where $\mathcal{A}$ is a slowly-varying envelope and the argument of the exponential is a rapidly-varying phase. It can be proven that this approximation is enough to derive Hawking radiation \cite{Visser2003}. Nonetheless, we can obtain a general solution of the wave equation \eqref{waveeq} if we define two new coordinates $u$ and $v$ as 
\begin{equation} \label{uvs}
u=t-\int^x \frac{\dd y}{c+V(y)}, \qquad v=t+\int^x \frac{\dd y}{c-V(y)},
\end{equation}
that solve the wave equation $ \partial_u \partial_v \phi=0$. Outside the region where $c=\abs{V}$, the solution of Eq. \eqref{waveeq} can be expressed as the sum of two arbitrary functions, one as an only $u$ function and the other as an only $v$ function:
\begin{equation} \label{izqder}
\phi = \phi^u \left( t-\int^x \frac{\dd y}{c+V(y)} \right) + \phi^v \left( t+\int^x \frac{\dd y}{c-V(y)} \right).
\end{equation}
Concerning a medium moving to the left, mode $\phi^u$ moves to the right and is known as the counter-propagating mode---it propagates against the fluid---while mode $\phi^v$ is the co-propagating mode---it propagates along with the fluid. Considering that Lagrangian \eqref{lag} is invariant to temporal translations, there are stationary mode solutions
\begin{equation}\label{tsym}
\phi(x,t)= e^{-i\omega t} \phi_\omega (x).
\end{equation}
Modes are then separated into counter-propagating and co-propagating parts $\phi=C_u \phi^u_\omega(x) +C_v \phi^v_\omega (x) $. Since the case under consideration does not include dispersive effects, such functional forms preserve its structure, where all wave components have speed $c$ measured from the moving medium. A local wavenumber $k(x)$ is defined such that $\phi^{u,v}_\omega (x) = \exp\left( i\int^x k^{u,v}_\omega (y) dy\right)$, thus if we use Eqs. \eqref{izqder} and \eqref{tsym} we get
\begin{equation}
k^u_\omega (x) = \frac{\omega}{c+V(x)}, \qquad k^v_\omega (x) =- \frac{\omega}{c-V(x)}.
\end{equation}
The same result can be obtained through the eikonal approximation \eqref{eikonal}. In general terms, the previous equations constitute the dispersion relation
\begin{equation}
[\omega-kV(x)]^2= c^2 k^2.
\end{equation}
We identify the Doppler-shifted frequency of the co-moving frame $\omega'=\omega-kV(x)$, which is of particular importance for the theory of optical analogues.

Assuming the existence of a horizon (a region where $V=-c$), counter-propagating modes in the moving medium will have a total zero velocity there. For simplicity, suppose that the horizon is located at $x=0$ and that the first derivative of $V(x)$ is not zero there. Since the $u$ coordinate becomes singular at this point, a linear approximation of the velocity profile is usually performed as
\begin{equation}\label{alfa}
V(x)\approx -c+\alpha x, \qquad \alpha = \left. \frac{\dd V(x)}{\dd x} \right|_{x=0}.
\end{equation}
If $\alpha >0$, the speed increases in the same direction as the moving medium, as in a black hole; if $\alpha <0$, the speed decreases in the direction of the flux, as in a white hole.

Under approximation \eqref{alfa} and near the horizon, the $u$ coordinate behaves as
\begin{equation}\label{usplit}
u\approx t-\frac{1}{\alpha}\ln\left( \frac{\alpha}{c} \abs{x}\right).
\end{equation}
This functional form establishes the presence of a horizon at $x=0$, and a separation of space into two regions delimited by $x=0$. Since there must exist a full set of modes supporting the existence of waves in all space, there must be two different counter-propagating modes for all frequencies in each region of space separated by the horizon, this is
\begin{equation}\label{left}
\phi^{u,\text{L}}_{\omega}(x,t) = \theta(-x)\frac{1}{\sqrt{4\pi c \omega}}e^{i\omega u}, \qquad 
\phi^{u,\text{R}}_{\omega}(x,t) = \theta(x)\frac{1}{\sqrt{4\pi c \omega}}e^{-i\omega u},
\end{equation}
where the change of sign in the exponent for the left $L$ and right $R$ regions is included due to the change of sign in the quantity $c+V$, assuring a positive norm for the modes.

Since we are in the dispersionless case, the speed of the field is a constant $c(r,t)=c$, so the value of $\alpha$ given in Eq. \eqref{alfa} reduces to the relation \eqref{kappa}. A set of ingoing waves from Eqs. \eqref{left} that corresponds to a single outgoing wave is identified as an out-mode. Similarly, a set of outgoing waves that correspond to a single ingoing wave defines an in-mode. A proper definition of these modes requires a finite time for the formation of the horizon---sometimes referred to as an apparent horizon---such that we can construct these modes when the horizon has not yet been formed. The in- and out-modes form two different sets of orthonormal $u$-modes that can be related through a Bogoliubov transformation: 
\begin{equation}\label{bogoliuv}
 \phi^{\text{out,R}}_{\omega} =\sum_j \alpha_j(\omega) \phi^{\text{in,L}}_{\omega,j}+\sum_j \beta_j(\omega) \phi^{\text{in,R}}_{\omega,j},
\end{equation}
where the coefficients $\alpha$ and $\beta$ fulfil the norm-conservation identity
\begin{equation}\label{normcon}
\sum_j[ \abs{\alpha_j(\omega)}^2-\abs{\beta_j(\omega)}^2]=1.
\end{equation}
The $\beta_j(\omega)$ term in the Bogoliubov transformation \eqref{bogoliuv} mixes positive- and negative-frequency modes as a result of a scattering process caused by the horizon.
As Hawking originally did, we can consider the vacuum state $|0\rangle$ defined as the absence of particles, which mathematically reads as
\begin{equation}
\hat{a}|0\rangle =0, \quad \forall \; \hat{a},
\end{equation}
where $\hat{a}$ is the annihilation operator that annihilates all modes in the basis. It is known from quantum field theory in curved spacetimes that when at least one $\beta$ coefficient in Eq. \eqref{bogoliuv} is non-zero, the definition of vacuum can change depending on the set of orthonormal modes taken as basis \cite{Birrell1982}. Then, in a scattering process the in-modes are different from the out-modes. The existence of negative modes in Eq. \eqref{bogoliuv} is essential for the creation of particles from vacuum, making norm conservation more akin to electric charge conservation (with positive and negative charged particles) than to mass conservation in classical mechanics (with only positive massive particles).

The expectation value of the number of outgoing particles in the in-vacuum has a bosonic thermal distribution, which is seen far from the horizon as a Planckian flux of outgoing particles \cite{Mukhanov2007}:
\begin{equation}
\langle N^{\text{out,R}}_{\omega} \rangle  = \frac{\delta (0)}
{\displaystyle\exp\left(\frac{2\pi \omega}{\alpha}\right)-1}.
\end{equation}
As in the astrophysical Hawking case---certainly as in the Unruh effect---in this generalized derivation the emitted radiation is thermal and its effective temperature depends on the acceleration of the moving medium \cite{Robertson2012jpb}, this is
\begin{equation}\label{HRA}
k_B T = \frac{\hbar\alpha}{2\pi}.
\end{equation}
This equation is the central piece of the general theory of Hawking radiation that can be applied to any system: astrophysical or analogue. Despite the fact that the dispersionless case was considered in this derivation, in analogue systems the presence of some kind of dispersion is unavoidable, which usually leads to a change in the spectrum, the Hawking spectrum may no longer be thermal. In the astrophysical case it is not clear if a dispersion mechanism exists, but if spacetime discretisation were true, dispersion effects could be introduced into Hawking radiation theory as in the analogue systems \cite{Leonhardt2010,Bermudez2019}.

The most efficient way of solving the Bogoliubov transformations for several in- and out-modes relies on the scattering matrix formalism. This method has been used to derive the Hawking spectrum for several systems, including the optical case that is a highly-dispersive system \cite{Robertson2014,Finazzi2014,Jacquet2018,Bermudez2016pra,MorenoRuiz2019}.

\section{Analogue Hawking radiation in optics}\label{secGAO}
\begin{figure}
	\centering\includegraphics[width=0.8\textwidth]{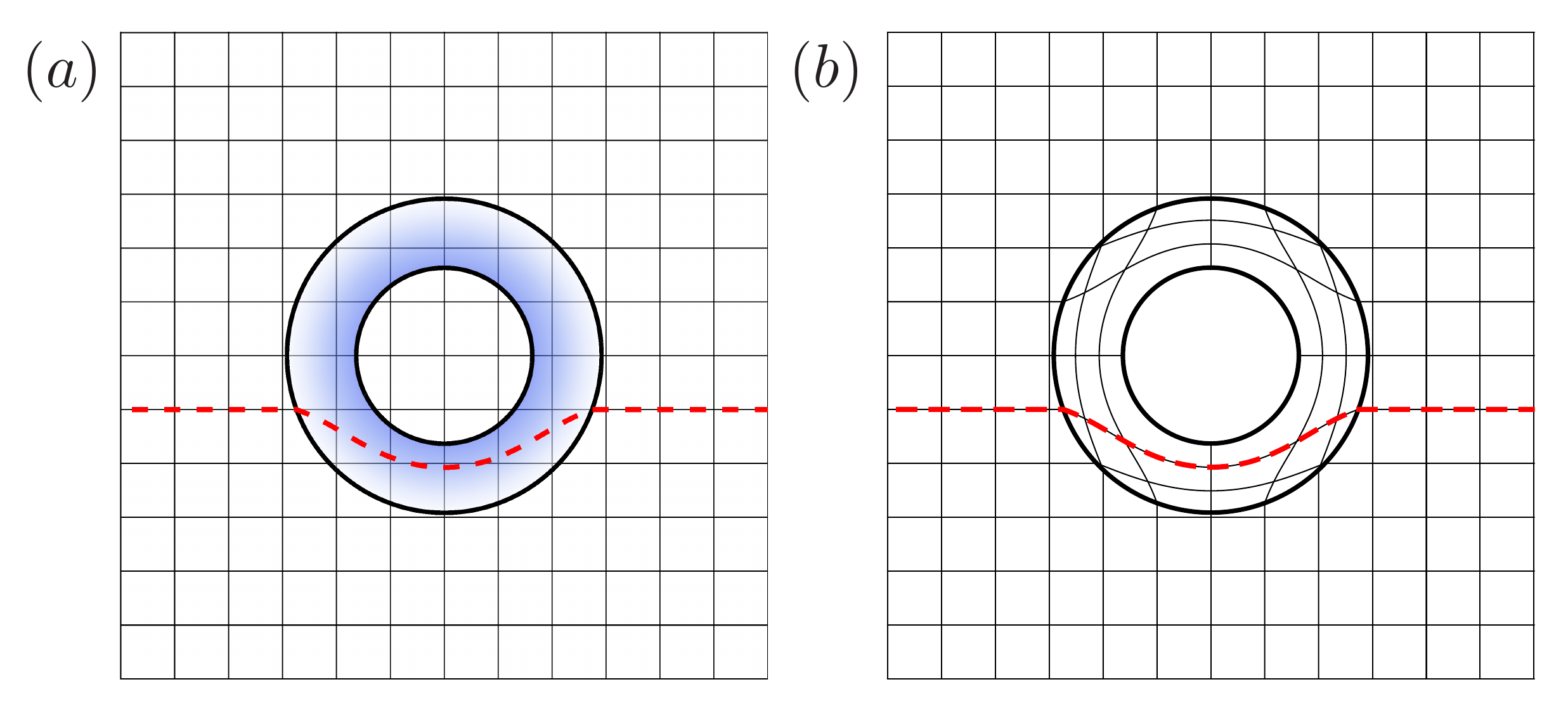}
	\caption{Transformation optics. In (a), flat space with non-isotropic media (blue). In (b), curved space (curved lines) in isotropic media. Also, an equivalent geodesic (red dashed) is shown in both spaces.}
	\label{fig_to}
\end{figure}

The behaviour of light depends on the propagating medium. Non-uniform media can bend light due to the change in the index of refraction caused, for example, by a hot air gradient as in a mirage. According to Einstein equations, a highly massive body can also bend light due to the spacetime curvature, as light propagates through geodesics. In the first case there is a flat geometry and non-isotropic media ($g^{ij}=\delta^{ij}$, $\epsilon^{ij}\neq\delta^{ij}$), whereas in the second case there is curved geometry and isotropic media ($g^{ij}\neq\delta^{ij}$, $\epsilon^{ij}=\delta^{ij}$), see Fig. \ref{fig_to} and Eq. \eqref{metric}. The equivalence between both systems was initially derived in Refs. \cite{Gordon1923,Plebanski1960} and eventually led to the establishment of the field of transformation optics \cite{Leonhardt2006,Pendry2006}. However, we need a moving horizon to pass from studying light bending in space to light bending in spacetime.

The practical realization of optical analogues is already on its way \cite{Philbin2008,Belgiorno2010,Petev2013,Drori2019}. The first obstacle for the creation of an optical black-hole lies in the necessity to have a medium moving at a speed close to the speed of light in the material. Initially, Ulf Leonhardt proposed a slow-light model \cite{Leonhardt2002}, where the group velocity of light is reduced by electromagnetically induced transparency. However, the modification of both phase and group velocities is necessary to obtain the analogue Hawking effect, for this reason slow light could not be implemented, as it only affects the group velocity \cite{Unruh2003}. Still, this work inspired a closer look into the relation between optics and general relativity. Similarly to Unruh's story, Leonhardt's interest in analogue gravity also arose from teaching a course, in his case of general relativity \cite{Leonhardt2020}.

\begin{figure}\centering
	\includegraphics[width=.47\textwidth]{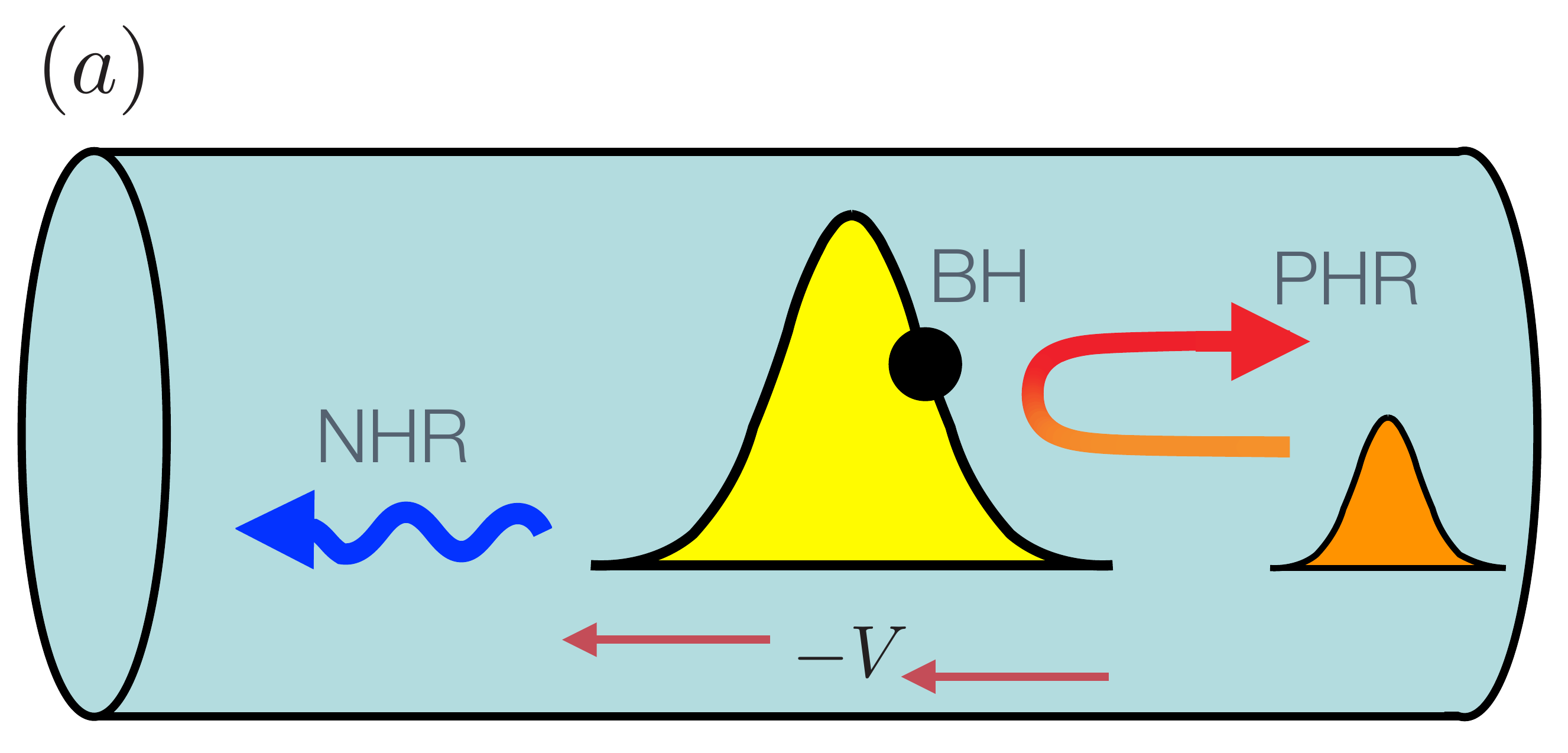}
	\includegraphics[width=.47\textwidth]{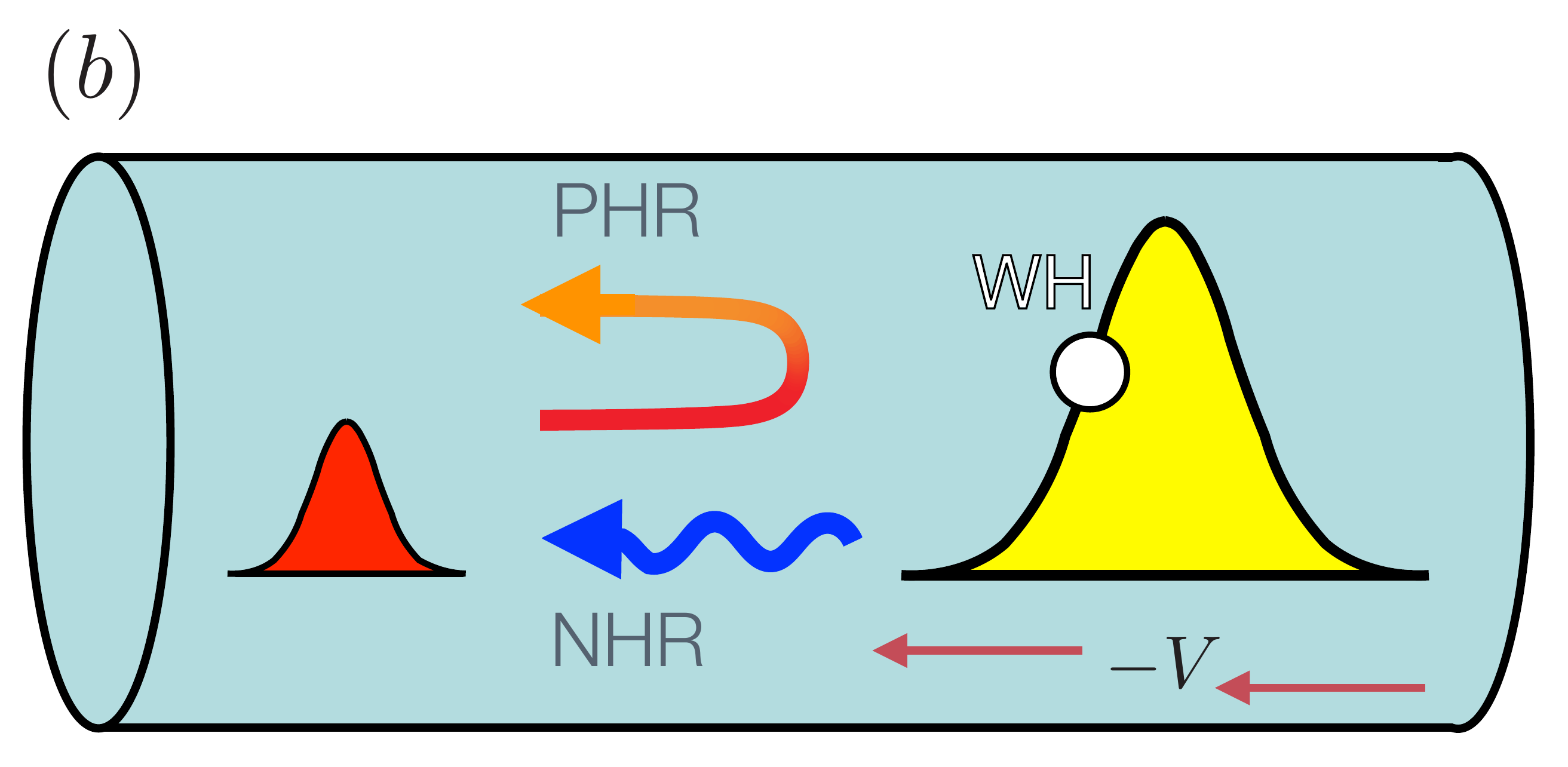}
	\caption{Pump pulse (yellow) is fixed at the centre and the fibre (green) moves to the left with speed $-V$ in the co-moving frame. In (a), the probe pulse (orange) moves to the left and interacts with a black hole (BH) horizon, in (b) the probe pulse (red) moves to the right and interacts with a white hole (WH) horizon. In both cases the probe produces two signals that correspond to the analogues of positive- (PHR) and negative- (NHR) frequency Hawking radiation.}
	\label{fig_fibre}
\end{figure}

With a change of perspective, moving an optical medium to a speed close to the speed of light is actually possible and a phenomenon that occurs all the time in optical telecommunications, where light pulses propagate through optical fibres. Acknowledging that the effective medium properties are the ones that determine the appearance of an event horizon, Thomas Philbin and colleagues proposed \cite{Philbin2008} to take advantage of the localized perturbation generated by the optical Kerr effect. This effect introduces a local change in the index of refraction proportional to the pulse intensity \cite{Agrawal2013}. Since this contribution travels along with the pulse, the pulse itself constitutes a moving medium: We call this the pump pulse. Note that no fibre material is being displaced, only light itself. In the reference frame of the pulse, the fibre material appears to be moving in the opposite direction. The index of refraction $n$ can be expressed as the sum of the intrinsic index of the medium $n_0$ plus a small contribution due to Kerr effect $\delta n$:
\begin{equation}
 n(\omega,t)=n_0(\omega) +\delta n(t).
\end{equation}
The speed of the effective medium is equal and opposite to the group velocity $V$ of the pump in the fibre. Another pulse---called probe---would propagate through a non-uniform index of refraction. Considering that in dispersive materials the speed of light depends on the frequency, we can work with probe light of a different frequency that moves slower [faster] than the pump, i.e., moving to the left [right] in the co-moving frame. This probe is then reversed by the interaction with the pump due to the change in the refractive index in a soft crash or blockage establishing a horizon at the point where the probe is slowed down to a stop or a point of no return. We create an artificial black [white] hole horizon depending on the relative motion of the probe with respect to the pump, as shown in Fig. \ref{fig_fibre}(a) [\ref{fig_fibre}(b)].

The effective geometry is established by the dielectric properties of the medium. These are obtained from the quantum field of light described by the vector potential $\hat{A}$ that generates the electromagnetic fields. For the first part of the calculations we ignore the $\omega$-dependence of $n$ and include it back in the dispersion relation \eqref{phi1}. Using the constitutive equations under a Lorentz transformation from an inertial frame---the laboratory frame---to a reference frame moving with the propagating pump---the co-moving frame---we get the relations \cite{Leonhardt2010}:
\begin{align}
 \hat{D}&= \epsilon_0 \gamma^2 \left[\epsilon (\hat{E}-V\hat{B})+\frac{V}{\mu}\left(\hat{B}- \frac{V}{c^2}\hat{E} \right) \right],\\
 \hat{H}&= \epsilon_0 \gamma^2 \left[\epsilon V (\hat{E}-V\hat{B})+\frac{V}{\mu}\left(\hat{B}+\frac{c^2}{\mu}\hat{E} \right) \right],
\end{align}
where $\gamma=1/\sqrt{1-V^2/c^2}$ is the relativistic factor and $\epsilon, \mu$ are the electric permittivity and magnetic permeability, respectively. We use the matrix $g^{\mu\nu}$ to introduce the analogy of the geometric background 
\begin{equation} \label{metric}
g^{\mu\nu} = \frac{1}{c^2-V^2}\begin{pmatrix} 
\epsilon c^2-\mu^{-1}V^2 & (\epsilon -\mu^{-1})cV   \\ 
(\epsilon -\mu^{-1})cV & \epsilon c^2-\mu^{-1}V^2 \
\end{pmatrix}.
\end{equation}
Thus we can establish the relations
\begin{equation}
\hat{D} =-\epsilon_0 c g^{0\nu}\partial_\nu \hat{A}, \quad  \hat{H} = -\epsilon_0 c^2 g^{1\nu}\partial_\nu \hat{A},
\end{equation}
where the Einstein sum convention has been used. The wave equation for light in a moving medium is similar to the one for a scalar field in curved spacetime, i.e.,
\begin{equation} \label{wel}
\partial_\mu g^{\mu \nu} \partial_\nu  \hat{A} = 0.
\end{equation}

In a similar way as the change of variables was previously used in Eq. \eqref{uvs}, we define a change of coordinates of the form
\begin{equation}
t_\pm =t-\int \frac{\dd x}{v_\pm},
\end{equation}
where the relativistic sum of velocities in the medium is done according to Einstein's velocity addition theorem
\begin{equation}\label{te}
v_\pm = \frac{\displaystyle V\pm \frac{c}{n}}{\displaystyle 1\pm \frac{V}{c n}}.
\end{equation}

The wave equation \eqref{wel} can be rewritten as
\begin{equation}\label{newwe}
4\frac{\Lambda}{c^2} \frac{\partial^2 \hat{A}}{\partial t_+ \partial t_-} = 0,
\end{equation}
where $\Lambda =n(c^2-V^2)/(c^2-n^2V^2)$ is a conformal factor. It can be appreciated that Eq. \eqref{newwe} leads to solutions \eqref{izqder}. A moving medium establishes an effective geometry described by Eq. \eqref{metric}, such that the horizon is formed when the velocity of the medium equals the speed of light in that medium:
\begin{equation}\label{hops}
 V =-\frac{c}{n},
\end{equation}
where the new coordinate $t_+$ contains a pole in the horizon, because $v_+$ is cancelled at this point and thus, as we did for the u-coordinate in Eq. \eqref{alfa}, we realize a first-order approximation of the form
\begin{equation}
v_+ =\alpha x, \qquad \alpha = \left. \frac{\dd v_+}{\dd x}\right|_{x=0}. 
\end{equation}

Using Eqs. \eqref{te} and \eqref{hops}, it can be showed that
\begin{equation}\label{alfa2}
\alpha = \frac{1}{1-n^{-2}} \left. \left(\frac{\dd V}{\dd x}-\frac{c}{n^2}\frac{\dd n}{\dd x} \right) \right|_{x=0}.
\end{equation}
Therefore, the corresponding Hawking temperature is given by 
\begin{equation}\label{HR}
k_B T = \frac{\hbar\alpha}{2\pi},
\end{equation}
i.e., the same as in Eq. \eqref{HRA}. These calculations were made in the co-moving frame. Since Hawking radiation is ultimately observed in the laboratory frame, it is necessary to perform a Lorentz transformation on the result \eqref{alfa2}. The coordinates for the co-moving frame are defined as
\begin{equation} \label{newcomov}
\tau = t-\frac{x}{V}, \qquad \zeta =\frac{x}{V},
\end{equation}
where $V$ appears as the frame moves with the pump. The retarded time or delay $\tau$ plays the role of distance, while the propagation time $\zeta$ plays the role of time. Then, $\alpha$ can be expressed as
\begin{equation}\label{surfgrav}
\alpha = \frac{1}{\delta n}\frac{\dd\delta n}{\dd\tau}.
\end{equation}
The Hawking temperature depends on the relative change of $\delta n$: a spectrum with considerable temperature could be produced with a very short pulse, making Hawking radiation an observable phenomenon in optical fibres \cite{Philbin2008}.

To include the dispersive properties of the medium modelled by its dispersion relation, we consider a first-order approximation that is valid around the horizon and allows us to keep Eq. \eqref{HR} with a varying $\alpha=\alpha(\tau)$ by evaluating Eq. \eqref{surfgrav}. The dispersion relation is usually expressed as a variation of wavenumber with frequency $k(\omega)=n(\omega)\omega/c$.

\begin{figure}
	\centering\includegraphics[width=.6\textwidth]{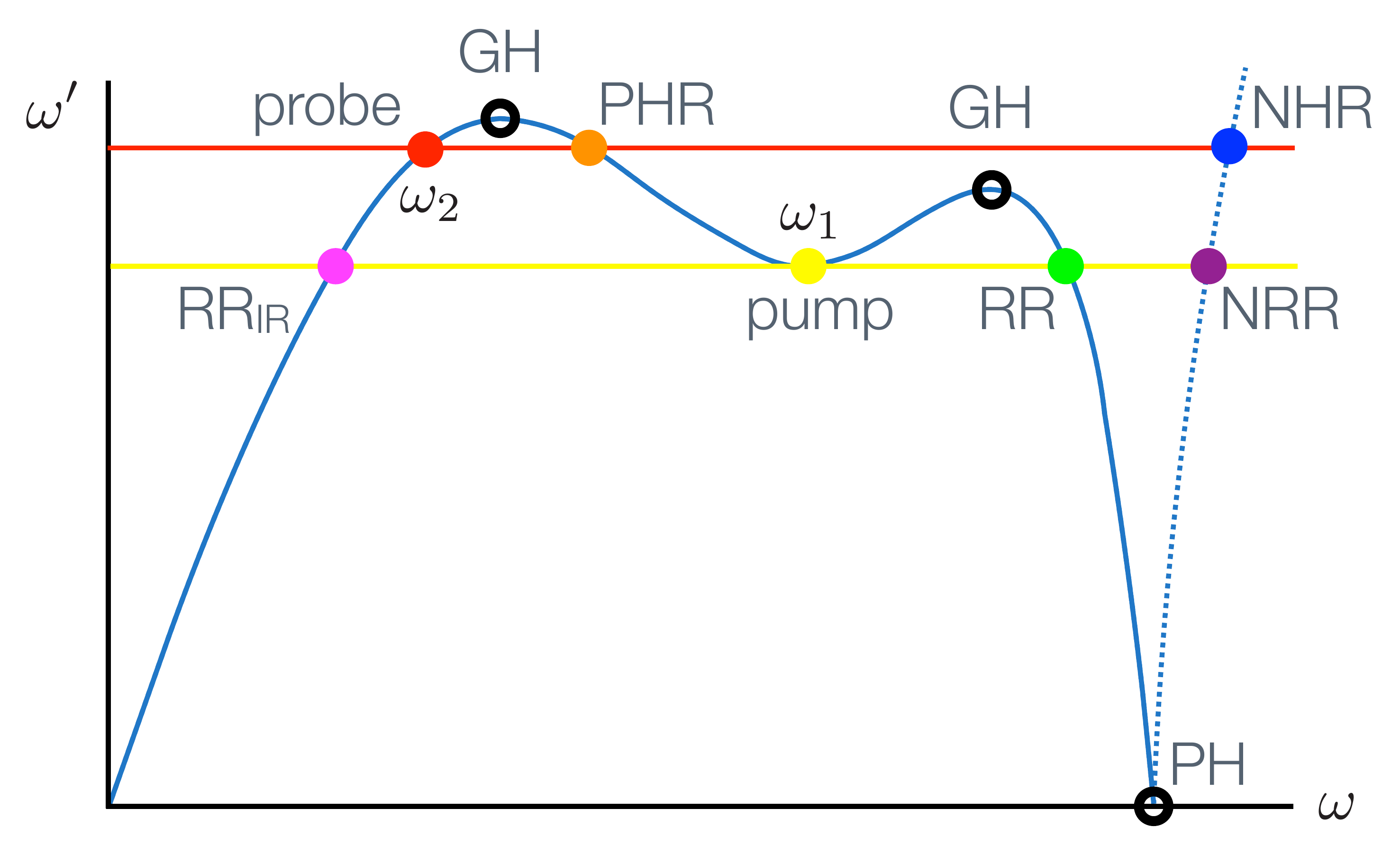}
	\caption{Dispersion relation in the co-moving frame $\omega'(\omega)$. Horizontal lines mark the conservation of $\omega'$ for the pump $\omega_1$ (yellow) and for the probe $\omega_2$ (red). The group- (GH) and phase- (PH) velocity horizons are also marked.}
	\label{fig_wprime}
\end{figure}

The phase of the field can gives us the conservation laws of the system. In the laboratory frame we have
\begin{equation}\label{phi1}
\varphi = \int (k \dd x -\omega \dd t),
\end{equation}
and under the change of coordinates \eqref{newcomov}, the phase in the co-moving frame is transformed to
\begin{equation}\label{phase-comov}
\varphi = -\int (\omega \dd\tau + \omega^\prime \dd\zeta), 
\end{equation}
where the co-moving frequency $\omega'$ is defined as
\begin{equation}\label{disprel}
\omega'(\omega) =\gamma \left( \omega -V k(\omega) \right) =\gamma \left(1-n(\omega)\frac{V}{c} \right) \omega.
\end{equation}
By comparing Eqs. \eqref{phi1} and \eqref{phase-comov}, we show that $-\omega$ and $\omega^\prime$ play the roles of $k$ and $\omega$, respectively \cite{Bermudez2016pra,GaonaReyes2017}. The functional form of $n(\omega)$ is modelled by considering the dispersive properties of the medium. It is convenient to use a photonic-crystal fibre (PCF) as the medium because it has a finite region of anomalous group-velocity dispersion (GVD), as the general shape depicted in Fig. \ref{fig_wprime}. The pump with laboratory frequency $\omega_1$ (yellow) sits in the anomalous GVD with positive curvature, and the probe with $\omega_2$ in the normal GVD region with negative curvature. In the high-frequency regime the dispersion relation goes to zero at the phase-velocity horizon and becomes negative. We plot the negative part of the dispersion as a dotted line to show it in the first quadrant.

The pump creates non-linear effects by itself, in particular, the conservation of its co-moving frequency creates new signals as resonant radiation (RR and RR${}_\textbf{IR}$), also known as Cherenkov radiation or dispersive waves \cite{Akhmediev1995}.

As we said, negative frequencies are essential for the derivation of Hawking radiation. In fact, the study of optical analogues led to the prediction of a new phase-matching condition for resonant radiation. This was previously overlooked because it needs a mixing of positive and negative frequencies. The new signal is called negative-frequency resonant radiation (NRR), it has been measured in several experiments \cite{Rubino2012,Drori2019,Mclenaghan2014} and explained by Fabio Biancalana and colleagues \cite{Conforti2013} using non-linear optics.

The need for both the group- and phase-velocity horizons becomes clear from the dispersion relation in Fig. \ref{fig_wprime}. The change of refractive index blocks the in-mode at the group horizon and undergoes a Hawking effect---redshifting and blueshifting of the in-modes occurs at the group horizon \cite{Choudhary2012}. The negative modes from the Bogoliubov transformation \eqref{bogoliuv} appear at the phase horizon, this is fundamental to create particles while still conserving the norm.

\begin{figure}
	\centering\includegraphics[width=.8\textwidth]{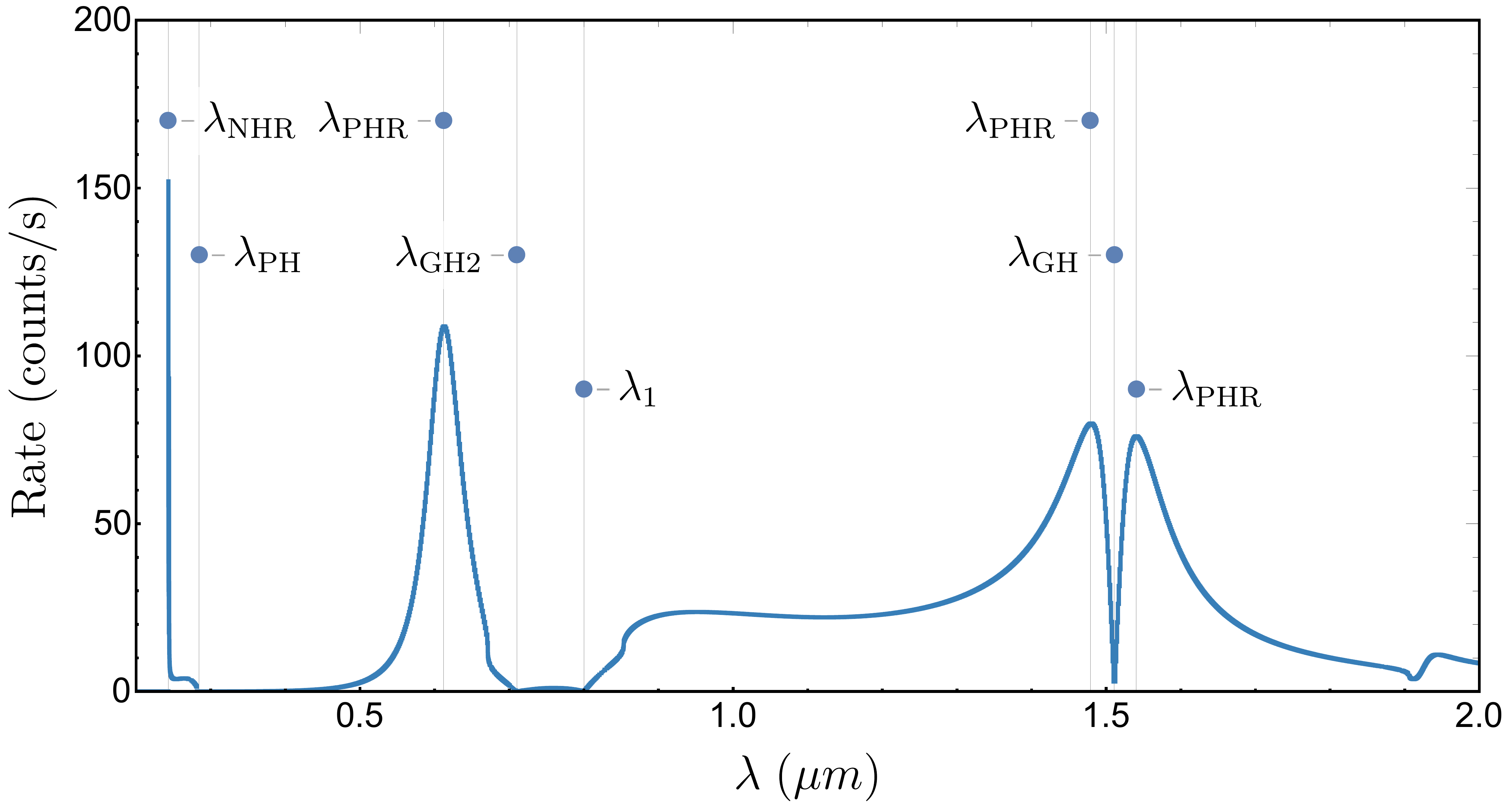}
	\caption{Spectrum from the spontaneous Hawking effect in a PCF from solving the mixing of Bogoliubov coefficients with the scattering matrix. The group and phase horizons are marked, as well as the frequencies where the peaks of Hawking emission are expected (PHR, NHR).}
	\label{fig_spectrum}
\end{figure}

The Bogoliubov transformation \eqref{bogoliuv} can be solved efficiently using the scattering matrix. In this way we obtain the Hawking spectrum, the intensity of Hawking radiation for different frequencies. Hawking radiation still exists even under dispersion, but its thermality is lost, or more precisely, the effective temperature depends now on the frequency. With this method \cite{Robertson2014,Bermudez2016pra}, we obtained a Hawking spectrum for the PCF used in experiment \cite{Drori2019}, shown in Fig. \ref{fig_spectrum}.

Note that in Fig. \ref{fig_spectrum} there is no probe $\lambda_2$ as this is a quantum Hawking effect that is caused directly from quantum vacuum. In a sense, our probe is the vacuum state with all wavelengths. All the signals from the phase horizon and redder are positive-frequency Hawking radiation (PHR)---this includes the peaks obtained from the first-order approximation marked as $\lambda_\text{PHR}$---and from the phase horizon and bluer are negative-frequency Hawking radiation (NHR). The complicated structure is partly due to the two group horizons in the realistic PCF dispersion.

We can now discuss the essential points for the Hawking effect to be realized in optics. In this case, the existence of a Lorentzian metric is given by Eq. \eqref{metric}. We recall the apparent horizon is established by a pulse travelling in the medium, causing a change in the local index of refraction by means of the Kerr effect. In dispersive systems the concept of apparent horizon depends on the frequency of the approaching modes. Each frequency mode will experience its own horizon because its group velocity will vary with frequency and each mode will be blocked at a different position (in $\tau$) of the pump: this is called a ``fuzzy horizon''. The value of the surface gravity $\alpha$ is given by \eqref{surfgrav}, it depends on $\tau$, being non-zero for all the frequencies that experience the horizon. As for the slow evolution, we recall that a soliton in an anomalous dispersion regime ($\beta_2<0$) will retain its shape since dispersive effects are counterbalanced by the nonlinear ones. The optical production of light through the Kerr effect has a response time of the electronic polarization of the material $10^{-15}$ s \cite{Boyd2008}, while the evolution of the pump in the fibre is on the order of $10^{-13}$ s.

\section{Stimulated Hawking radiation in optics}
Hawking radiation was originally obtained with quantum vacuum as the input state. However, the Hawking effect still occurs if we replace the vacuum with a different state, this is called stimulated radiation. The stronger stimulated effect is obtained if we replace vacuum with a classical state. This is classical stimulated Hawking radiation and is the usual effect studied in optical analogues. In this case we use a probe pulse or a continuous wave (CW) along with the pump pulse to generate stimulated Hawking radiation, which is easier to detect. All optical experiments so far have measured the classical stimulated effect, in 2008 it was done with a CW \cite{Philbin2008} and the latest in 2019 with a pulse \cite{Drori2019}.

We study the pulse dynamics inside the fibre to explain the production of new signals from the point of view of optics. The most used model is the non-linear Schr\"odinger equation (NLSE):
\begin{equation}
i\frac{\partial E}{\partial z}-\frac{|\beta_2|}{2}\frac{\partial^2 E}{\partial t^2}+\gamma|E|^2E=0,
\end{equation}
that describes the electric part $E(z,t)$ of the optical field based on the second-order dispersion $\beta_2$ and the non-linear coefficient $\gamma$. To increase the intensity of the Hawking signal we need to increase the surface gravity $\alpha$ and from Eq. \eqref{surfgrav} this means a fast changing $\delta n$. Therefore, we should use the shortest available optical pulses, with duration of around 7 fs. However, these pulses cannot be described by the NLSE, but with a different version with less approximations \cite{Couairon2011} called unidirectional pulse propagation equation (UPPE):
\begin{equation}\label{uppe}
i\frac{\partial E_\omega} {\partial z}+\beta(\omega)E_\omega+ \frac{\omega}{2cn(\omega)}P_{\text{NL},\omega}=0,
\end{equation}
where $E_\omega$ is the Fourier transform of the electric field, $\beta(\omega)=k(\omega)$ is the full dispersion relation, and $P_\text{NL}=\chi^{(3)}E^3/8$ is the non-linear polarization.

Dissipation in optical systems can be included by an extra term in the evolution equation, sometimes called generalized nonlinear Schr\"odinger equation (GNLSE). However, the dynamics of the optical Hawking effect are much faster than such dissipation and neglecting it is a very good approximation \cite{Agrawal2013}. On the other hand, dispersion produces soliton fission and the emission of dispersive waves at a similar length scale as the Hawking effect. For example, for the PCF used dissipation should be considered at 100m, dispersion at 10mm, and the Hawking effect is produced in 10mm \cite{Dudley2006,Amiranashvili2019}. It was shown in the numerical solutions and the experiment in Ref. \cite{Drori2019} that the Hawking effect is robust enough to survive these strong dispersive effects.

Equation \eqref{uppe} describes the classical evolution of the ultra-short optical pulses inside an optical fibre, as our starting equations are Maxwell equations. This framework can only describe the emission of the stimulated Hawking radiation, not the spontaneous one. We set the pump and probe signals as part of the electric field $E$ and we obtain the signal from the stimulated Hawking effect.

As the same frequency conservation is fulfilled in both the stimulated and the spontaneous cases, we obtain Hawking signals in the correct frequencies described by the conservation of co-moving frequency of the probe $\omega'(\omega_2)$, but these signals are created from a classical frequency-shifting of the probe energy and not directly from vacuum energy \cite{AgueroSantacruz2020}.

The existence of the probe pulse at frequency $\omega_2$ causes a different co-moving frequency conservation, as seen in Fig. \ref{fig_wprime}. The probe (red) creates signals at two frequencies, the positive- (PHR, orange) and negative- (NHR, blue) frequency Hawking radiation. This situation corresponds to the black-hole horizon shown in Fig. \ref{fig_fibre}(a); for the white-hole horizon, probe and PHR should be interchanged.

\begin{figure}
	\centering\includegraphics[width=.8\textwidth]{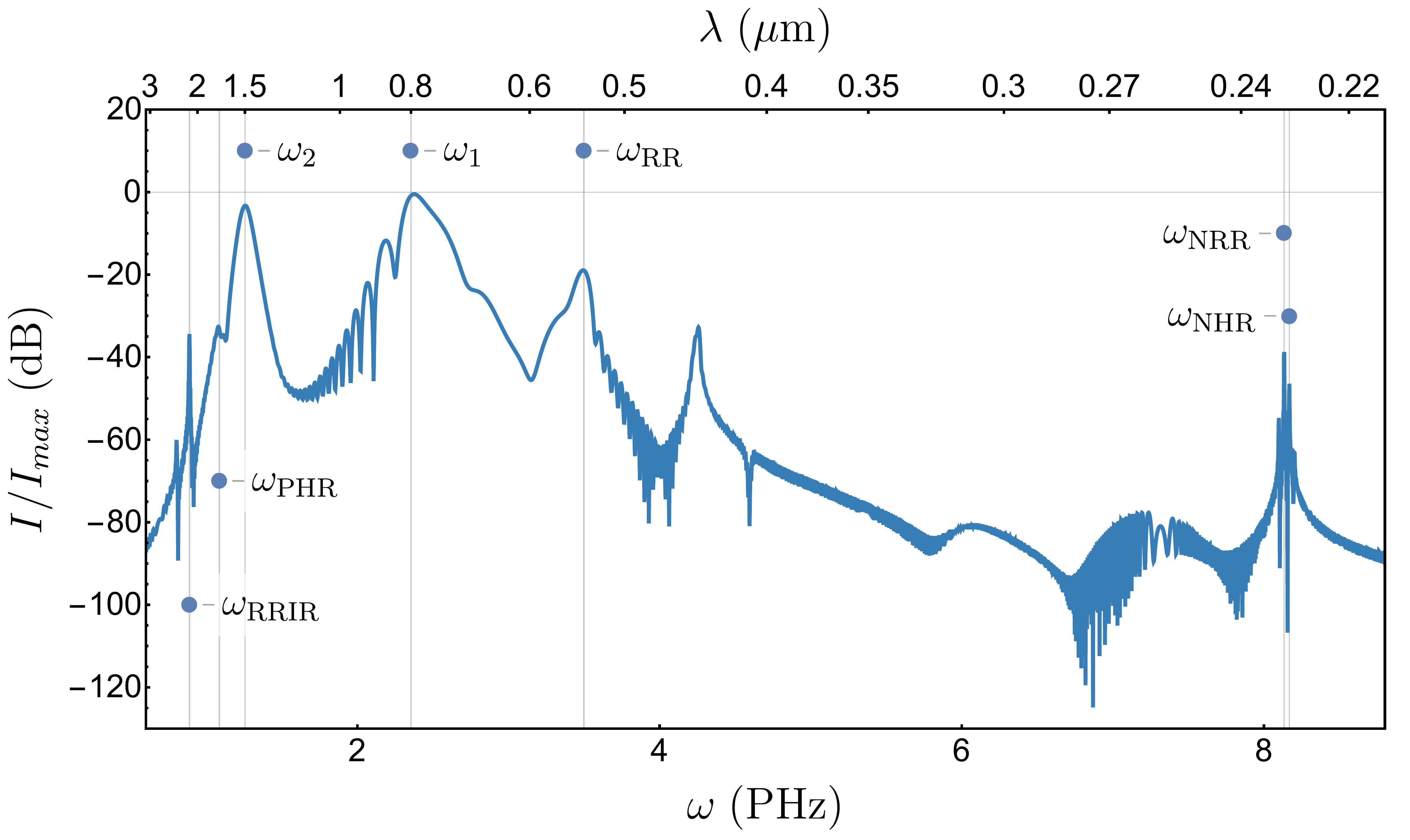}
	\caption{Spectrum of pump $\omega_1$ - probe $\omega_2$ interaction after 7mm of propagation in a PCF fibre obtained through numerical solution of the UPPE \eqref{uppe}. The stimulated Hawking effect is shown as NHR and PHR and the resonant radiation as RRIR, RR, and NRR.}
	\label{fig_dynamics}
\end{figure}

By working out the pulse dynamics, we can obtain the output spectrum that includes new stimulated signals at both the PHR and NHR frequencies. An example of these spectra is shown in Fig. \ref{fig_dynamics} by solving Eq. \eqref{uppe} for a 7fs, 800nm, 50mW pump and a 50fs, 1500nm, 5mW probe propagating through 7mm of PCF \cite{Drori2019}.

\section{Conclusion}
We studied the emission of Hawking radiation in its original derivation in the astrophysical case and we argued for the generalization of such concept to other systems. These are called analogue systems and the emission is called analogue Hawking radiation. We mentioned examples that have experimental evidence, and focus only in light pulses in optical fibres.

We presented a derivation of Hawking radiation in a general setting that works for any non-dispersive system, in particular, this includes the original derivation in astrophysics but also in any analogue system under a non-dispersive approximation. In this way it is clear that the Hawking effect is more general than previously thought. We extended this derivation to dispersive systems using a linear approximation.

Then, we focused on the optical analogues: We explained the main ideas and established that the conditions for the emission of analogue Hawking radiation are fulfilled: apparent horizon, surface gravity, and slow evolution.

We studied the expected signals in the quantum or spontaneous case (Fig. \ref{fig_spectrum}) and in the classical or stimulated one (Fig. \ref{fig_dynamics}). Each with its own way of calculating the resulting spectrum for the Hawking emission. For the spontaneous radiation we used the scattering matrix to solve the mixing of Bogoliubov coefficients \cite{Robertson2014,Bermudez2016pra,Jacquet2018}. For the stimulated one, the pulse dynamics gives the classical emission in non-linear optics \cite{Drori2019,Rubino2012sr}.

\begin{figure}
\centering\includegraphics[width=.9\textwidth]{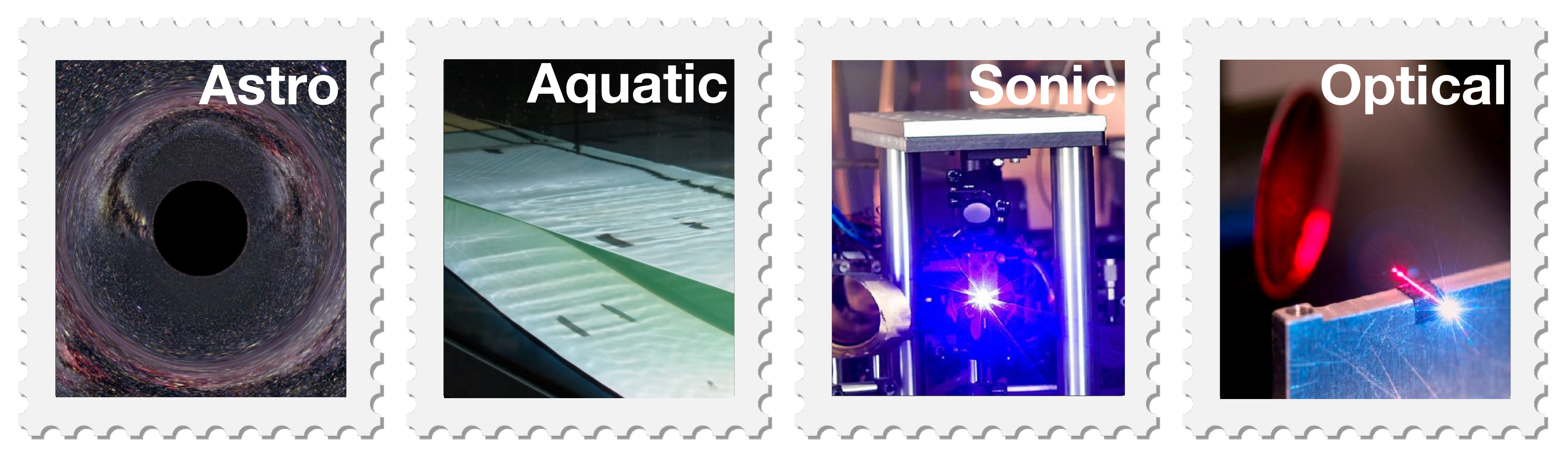}
\caption{Instead of collecting ``stamps'' or several instances of phenomena similar to Hawking radiation, we should generalize the concept and call them all Hawking radiation. Images: [Astro] Ute Kraus, Universit\"at Hildesheim \cite{Kraus2020}; [Aquatic] Germain Rousseaux, Institut Pprime \cite{Euve2016}, [Sonic] Jeff Steinhauer, Technion \cite{Munoz2019}; [Optical] Yuval Rosenberg, Weizmann Institute of Science \cite{Drori2019}.}
\label{fig_stamps}
\end{figure}

We would like to finish with a comment on the current state of the field of analogue gravity in front of the scientific community. ``All science is either physics or stamp collecting'', this phrase is usually attributed to Ernest Rutherford. It is a controversial phrase, specially if it is used by physicists. However, we would like to use it to counter an argument usually made by physicists against the use of the name Hawking radiation to describe the so-called analogue Hawking radiation, i.e., a Hawking-type effect in analogue systems, see Fig. \ref{fig_stamps}. The naming of physical effects is important to emphasize if they are related at all and how closely. In this case, instead of collecting several related facts or ``stamps'' and giving a different name to each one---astrophysical Hawking radiation, aquatic Hawking radiation, sonic Hawking radiation, optical Hawking radiation---we should generalize the concept of Hawking radiation. Yes, it was discovered originally in astrophysics, but it is realized in the same way in all analogue systems, as it was shown in the general derivation in Section \ref{general}. This phenomenon is nothing more than Hawking radiation.

\section*{Acknowledgements}
The authors would like to acknowledge the valuable discussions with Maxime Jacquet, Yuval Rosenberg, and Ulf Leonhardt. Also, we are grateful to Friderich K\"onig, Silke Weinfurtner, and Maxime Jacquet as the organisers of the scientific meeting on ``The next generation of analogue systems'' at the Royal Society on December 2019 that sparked this discussion. Funding provided by Fondo SEP-Cinvestav 2018-60 (Mexico).


\end{document}